\documentclass[a4paper,11pt]{article}
\usepackage{aaskaiid}
\usepackage{orcidlink}

\title{Probing Magnetic Fields In and Around Galaxies Near and Far}
\ShortTitle{Probing Magnetic fields in and around galaxies near and far}
\ShortName{Mao et al.} 

\author[1]{S.~A.~Mao\,\orcidlink{0000-0001-8906-7866}}
\author[1]{Rainer~Beck\,\orcidlink{0009-0001-8154-3562}}
\author[2,1]{Aritra~Basu\,\orcidlink{0000-0003-2030-3394}}

\author[3]{Lerato~Baidoo\,\orcidlink{0000-0003-0520-0696}}
\author[4,5,6]{Andrea~Bracco\,\orcidlink{0000-0003-0932-3140}}
\author[7]{Ralf-Jürgen~Dettmar\,\orcidlink{0000-0001-8206-5956}}
\author[8]{Volker~Heesen\,\orcidlink{0000-0002-2082-407X}}
\author[9]{Cathy~Horellou\,\orcidlink{0000-0002-3533-8584}}
\author[10]{Timea~O.~Kovacs,\orcidlink{0000-0001-6649-8559}}
\author[11]{Kohei~Kurahara,\orcidlink{0000-0003-2955-1239}}
\author[12]{Yu-Qing~Lou}
\author[1]{Yik~Ki~Ma\,\orcidlink{0000-0003-0742-2006}}
\author[13]{Rikuto~Omae\,\orcidlink{0000-0002-7259-1706}}
\author[14]{Rosita~Paladino\,\orcidlink{0000-0001-9143-6026}}
\author[15]{Amit~Seta\,\orcidlink{0000-0001-9708-0286}}
\author[16,10,1]{Fatemeh~Tabatabaei\,\orcidlink{0000-0002-0377-0970}}
\author[17]{Theresa~Wiegert\,\orcidlink{0000-0002-3502-4833}}

\affiliation[1]{Max-Planck-Institut f\"ur Radioastronomie, Auf dem H\"ugel 69, 53121 Bonn, Germany}
\emailAdd{mao@mpifr-bonn.mpg.de}
\affiliation[2]{Th\"uringer Landessternwarte, Sternwarte 5, 07778 Tautenburg, Germany}
\affiliation[3]{Dunlap Institute for Astronomy and Astrophysics, University of Toronto, 50 St. George Street, Toronto, M5S 3H4, ON, Canada}

\affiliation[4]{LUX, Observatoire de Paris, Université PSL, Sorbonne Université, CNRS, 75014 Paris, France}
\affiliation[5]{INAF – Osservatorio Astrofisico di Arcetri, Largo E. Fermi 5, 50125 Firenze, Italy}
\affiliation[6]{Laboratoire de Physique de l'Ecole Normale Sup\'erieure, ENS, Universit\'e PSL, CNRS, Sorbonne Universit\'e, Universit\'e de Paris, F-75005 Paris, France}
\affiliation[7]{Ruhr University Bochum, Faculty of Physics and Astronomy, Astronomical Institute (AIRUB), Universit\"atsstrasse 150, 44780 Bochum, Germany}
\affiliation[8]{University of Hamburg, Hamburger Sternwarte, Gojenbergsweg 112, 21029 Hamburg, Germany}
\affiliation[9]{Department of Space, Earth and Environment, Chalmers University of Technology, Onsala Space Observatory, 43992 Onsala, Sweden}
\affiliation[10]{Max-Planck-Institut f\"ur Astronomie, K\"onigstuhl 17, D-69117 Heidelberg, Germany }
\affiliation[11]{Kobayashi-Maskawa Institute for the Origin of Particles and the Universe, Nagoya University, Furo-cho, Chikusa-ku, Nagoya, Aichi 464-8601, Japan}
\affiliation[12]{Dept. of Physics and Tsinghua Center for Astrophys., Tsinghua Univ., Beijing, 100084, China}
\affiliation[13]{Department of Social Design Engineering, National Institute of Technology, Kochi College, 200-1 Monobe, Nankoku, Kochi 783-850, Japan}
\affiliation[14]{INAF Istituto di Radioastronomia, Via P. Gobetti 101, 40129 Bologna, Italy}
\affiliation[15]{Research School of Astronomy and Astrophysics, Australian National University, Canberra, ACT 2611, Australia}
\affiliation[16]{School of Astronomy, Institute for Research in Fundamental Sciences (IPM), P.O. Box 1956836613, Tehran, Iran}
\affiliation[17]{Instituto de Astrof\'isica de Andaluc\'ia, Glorieta de la Astronom\'ia S/N 18008 Granada, Spain}

\abstract{
In order to understand the magnetization of galaxies and the role of magnetic fields in feedback processes that govern star formation and galaxy evolution, it is essential to have a comprehensive census of magnetic fields in and around galaxies from the nearby Universe to high redshifts. 

For nearby galaxies, the geometry and strength of their galactic magnetic fields will be determined via the diffuse polarized synchrotron emission across multiple SKA bands, while deep rotation measure (RM) grids extending out to large galactocentric radii will provide constraints on the geometry and strength of the circumgalactic medium (CGM) magnetic fields. Beyond the local Universe, a promising approach to track the buildup of galactic magnetic fields is to utilize polarized background radio sources which are strongly gravitationally lensed by a distant galaxy lying in between. The differential Faraday rotation between lensed images serves as a direct tracer of the magnetized interstellar medium (ISM) in the lensing galaxy. A large sample of background radio sources lensed by foreground galaxies across a wide range of redshifts and lensing galaxy mass and type will allow us to trace how galactic magnetic fields evolve as a function of redshift and directly constrain galactic dynamo theories. Statistical rotation measure studies of back-illuminated galaxies will allow us to probe the magnetization of the CGM out to high redshifts. 

In this chapter, we outline the science goals, strategies, techniques, and observational requirements with SKA AA4 for (1) a homogeneous polarimetric survey of nearby galaxies -- mapping both the diffuse polarized emission as well as producing a dense RM grid within the virial radius; (2) a survey of the interstellar magnetic fields in distant galaxies targeting strong lensing systems with polarized lensed quasars, as well as a general statistical back-illumination survey to probe the redshift evolution of magnetic fields in the CGM. These proposed observations will serve as a major step towards understanding the co-evolution of galaxies and their magnetic fields over cosmic time and provide constraints on galactic dynamo theories.
}

\begin{document}
\newcommand{\actaa}{Acta Astron.} 
\newcommand{\araa}{ARA\&A} 
\newcommand{\aar}{A\&ARv} 
\newcommand{\aapr}{A\&ARv} 
\newcommand{\ab}{Astrobiol.} 
\newcommand{\aj}{AJ} 
\newcommand{\apj}{ApJ} 
\newcommand{\apjl}{ApJL} 
\newcommand{\apjs}{ApJSS} 
\newcommand{\ao}{Appl. Opt.} 
\newcommand{\apss}{Astro. \& Space Sci.} 
\newcommand{\aap}{A\&A} 
\newcommand{\aaps}{A\&AS.} 
\newcommand{\baas}{Bull. Am. Astron. Soc.} 
\newcommand{\caa}{Chinese A\&A} 
\newcommand{\cjaa}{Chinese J. A\&A} 
\newcommand{\cqg}{Class. Quantum Gravity} 
\newcommand{\gal}{Galaxies} 
\newcommand{\gca}{Geo. Cosmo. Acta} 
\newcommand{\icarus}{Icarus} 
\newcommand{\jcap}{JCAP} 
\newcommand{\jgr}{J. Geophys. Res.} 
\newcommand{\jgrp}{J. Geophys. Res. Planets} 
\newcommand{\jqsrt}{J. Quant. Spectrosc. Radiat. Transf.} 
\newcommand{\memsai}{Mem. SAIt} 
\newcommand{\mnras}{MNRAS} 
\newcommand{\nat}{Nature} 
\newcommand{\nastro}{Nat. Astron.} 
\newcommand{\ncomms}{Nat. Commun.} 
\newcommand{\nphys}{Nat. Phys.} 
\newcommand{\na}{New Astron.} 
\newcommand{\nar}{New Astron. Rev.} 
\newcommand{\physrep}{Phys. Rep.} 
\newcommand{\pra}{Phys. Rev. A} 
\newcommand{\prb}{Phys. Rev. B} 
\newcommand{\prc}{Phys. Rev. C} 
\newcommand{\prd}{Phys. Rev. D} 
\newcommand{\pre}{Phys. Rev. E} 
\newcommand{\prx}{Phys. Rev. X} 
\newcommand{\prl}{Phys. Rev. Let.} 
\newcommand{\psj}{Planet. Sci. J.} 
\newcommand{\planss}{Planet. Space Sci.} 
\newcommand{\pnas}{Proc. Natl Acad. Sci. USA} 
\newcommand{\procspie}{Proc. SPIE} 
\newcommand{\pasa}{PASA} 
\newcommand{\pasj}{PASJ} 
\newcommand{\pasp}{PASP} 
\newcommand{\rmxaa}{RMXAA} 
\newcommand{\sci}{Science} 
\newcommand{\sciadv}{Sci. Adv.} 
\newcommand{\solphys}{Sol. Phys.} 
\newcommand{\sovast}{Soviet Ast.} 
\newcommand{\ssr}{Space Sci. Rev.} 
\newcommand{\uni}{Universe} 

\maketitle

\section{Introduction}

From the plasma physics point of view, galaxies are prime laboratories for testing all flavors of dynamo theories because they are the only astrophysical systems in which the dynamo-active region is transparent and can be directly observed. The small-scale (or the fluctuation) dynamo, which just requires turbulence, is thought to dominate early on in young galaxies, amplifying the magnetic fields to considerable strengths. When the galactic disk settles into equilibrium, the large-scale (or the mean-field or $\alpha$-$\Omega$) dynamo begins to operate effectively, amplifying and organizing magnetic fields on large scales ($>$ kpc) from the combined effects of helical turbulence and differential rotation \citep{ruzmaikin1988,rincon2021,ss2021}. Unfortunately, this amplification and organization of magnetic fields in and around galaxies is not at all well constrained observationally. Only a handful of rigorous comparisons between the predictions of dynamo theory and observed field properties exist for nearby galaxies \citep{vaneck2015,beck2019}. 

While mean-field dynamo predictions are broadly consistent with basic properties of the observed large-scale galactic magnetic fields in the present day Universe, such comparisons at z$>$0 are not currently possible. There remain critical challenges and unresolved mysteries that can only be effectively addressed with new, sensitive and high-angular resolution radio polarization data from the SKA.

The magnetic field is a major component of the ISM and CGM that can influence star formation and galaxy evolution. Indeed, the presence of magnetic fields (and cosmic rays) is capable of strongly altering basic galaxy properties (e.g., the dominance of the galactic disk, mass of the central black hole) as well as the property and structure of the CGM such as the density and metal enrichment \citep{buck2020,vandevoort2021,ps2013}. 
Non-thermal pressure from magnetic fields and cosmic rays is capable of driving energetic winds that are ubiquitous in galaxies at high redshift \citep{hanasz2013,pakmor2016}. These winds magnetize the CGM and in turn would affect accretion / infall of gas onto galaxies at later times \citep[e.g][]{betti2019,berlok2019,jung2023}. Hence, understanding galaxy evolution is not possible without a much improved understanding of the evolution of magnetic fields in and around galaxies.

It has now become clear that magnetic fields should be studied in a broad context as basic galactic properties such as star-formation rate, galactic rotation and disk-halo connection shape galactic magnetic fields and vice versa. In that context, non-magnetic, non-polarization observations and measurements of the multiphase ISM are also crucial, particularly when building galaxy-specific dynamo models and their field-structure predictions \citep[e.g.][]{chamandy2024}. 

In this chapter, we build upon and update topics covered in the \cite{beck2015} AASKA science book chapter on magnetic fields in external galaxies. We focus mostly on the large-scale (kpc-scale) component of the galactic magnetic field and refer readers to \citep{Ma01.2026.SKA} in the same volume for studies of small-scale ($<$ 100 pc) magnetic fields with the SKA. Similarly, we refer readers to \citep{Sun01.2026.SKA} in the same volume for studies of the Milky Way magnetic field with the SKA. For broader and historical context of the general cosmic magnetism science case, we refer readers to \cite{gaensler2004,mao2020,heald2020}. 

\section{Theoretical and simulation expectations}

In the ISM, the small-scale dynamo seeds the mean-field dynamo by amplifying fields to $\mu$G strengths on relatively short time scales (turbulent eddy turnover timescales, $\approx$ 10$^7$ years) in young galaxies \citep{bs2005,seta2020,be2023}. Then, the mean-field dynamo organizes the field to coherent structures and amplifies its strength to $\mu$G level. In the disk, the fastest growing solution to the mean-field dynamo equation is axisymmetric with even parity across the mid-plane, while the field in the thick disk (halo)\footnotemark[1]\footnotetext[1]{Throughout this chapter, we refer to vertical heights up to several kpc as the thick disk, up to $\sim$ 10 kpc as the halo and beyond that as the CGM.} has odd parity across the mid-plane \footnotemark[2]\footnotetext[2]{Even parity disk field and odd parity halo field can co-exist in the presence of a galactic wind \citep{moss2010}, or one might enslave the other \citep{ms2008}.}. The coherent magnetic fields generated by the large-scale dynamo has spiral geometry with pitch angles on the order of 10$^{\circ}$ (trailing) that decreases at larger galactocentric radii for a flared disk. 

Early works on dynamo time scales in different types of galaxies have demonstrated, for example, that at z=0.5, a Milky Way-like galaxy would have developed coherent magnetic fields on kpc scales \citep{arshakian2009}. More sophisticated predictions of galactic magnetic field evolution have been obtained by coupling semi-analytical galaxy formation models with non-linear mean-field dynamo theory \citep{rodrigues2015,rodrigues2019} on a large sample of spiral galaxies. These studies showed that large-scale magnetic fields could be amplified to $\mu$G field strength between redshift of 2 and 3, depending on the galaxy mass. This work also produced important predictions of the large-scale magnetic field statistics, including the fraction of galaxies hosting large-scale magnetic fields at redshift zero\footnotemark[3]\footnotetext[3] {We are not yet able to test this at z=0 due to the lack of an unbiased large sample of galaxies with magnetic field measurements.}.

In the CGM (that is, the medium that lies beyond the ISM, but within the virial radius), cosmological magnetohydrodynamic (MHD) zoom simulations of Milky Way-like galaxies have shown that  magnetization occurs in two distinct phases. At early times, magnetization is driven by outflows that also produce metal enrichment. At z$<$1, an in situ turbulent dynamo further amplifies the CGM magnetic field, reaching a field strength of $\sim$ 0.1 $\mu$G that permeates the CGM out to the virial radius by z=0. In general, the CGM field is not expected to be well ordered, but may be aligned by coherent outflow velocity structures at low redshifts \citep{pakmor2020}. 

\section{State of the Art in Nearby Galaxies}

\begin{figure}[h]
    \centering

    \begin{minipage}{0.48\textwidth}
        \centering
        \includegraphics[width=\linewidth]{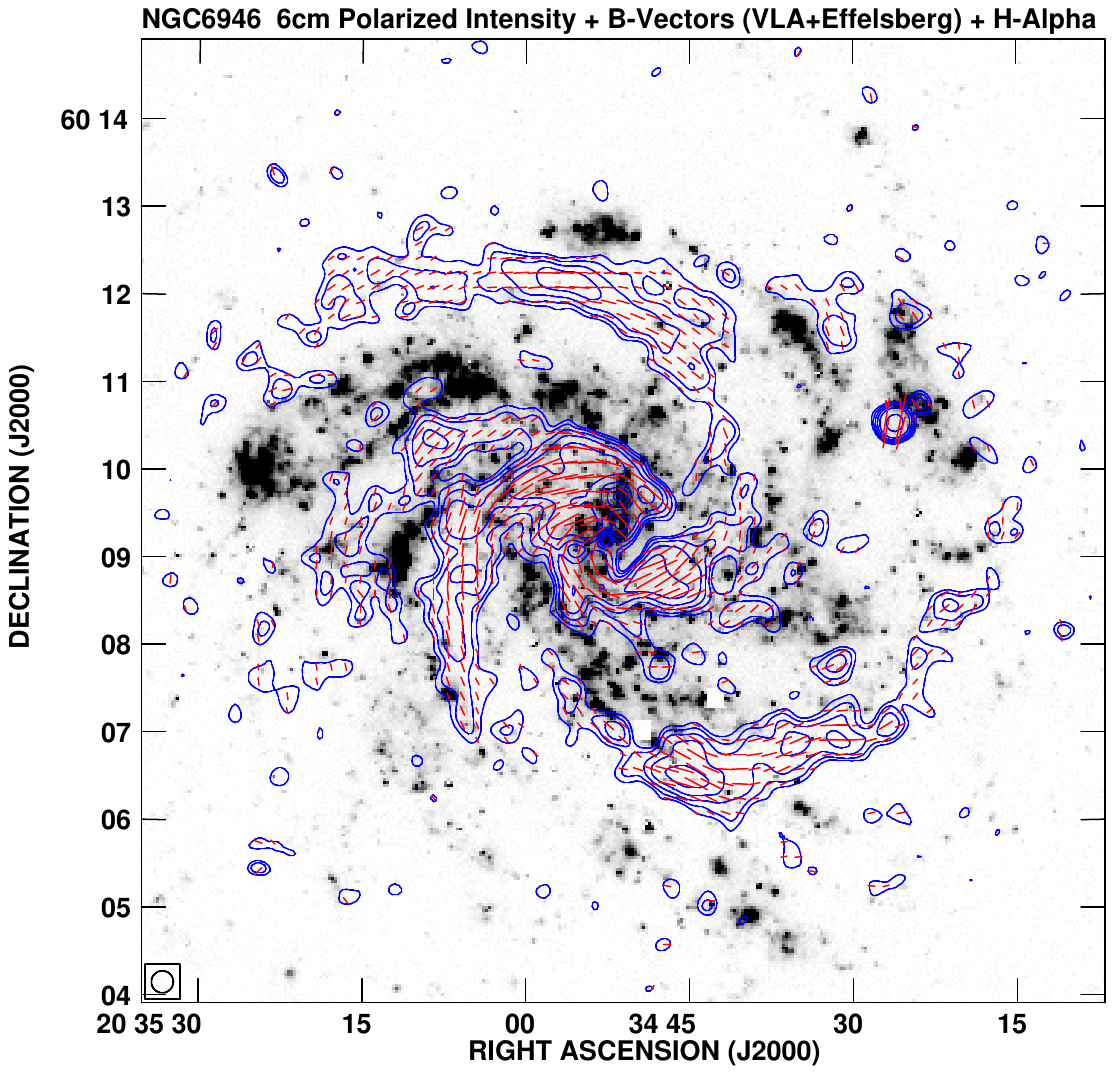}
    \end{minipage}
    \hfill
    \begin{minipage}{0.48\textwidth}
        \centering
        \includegraphics[width=\linewidth]{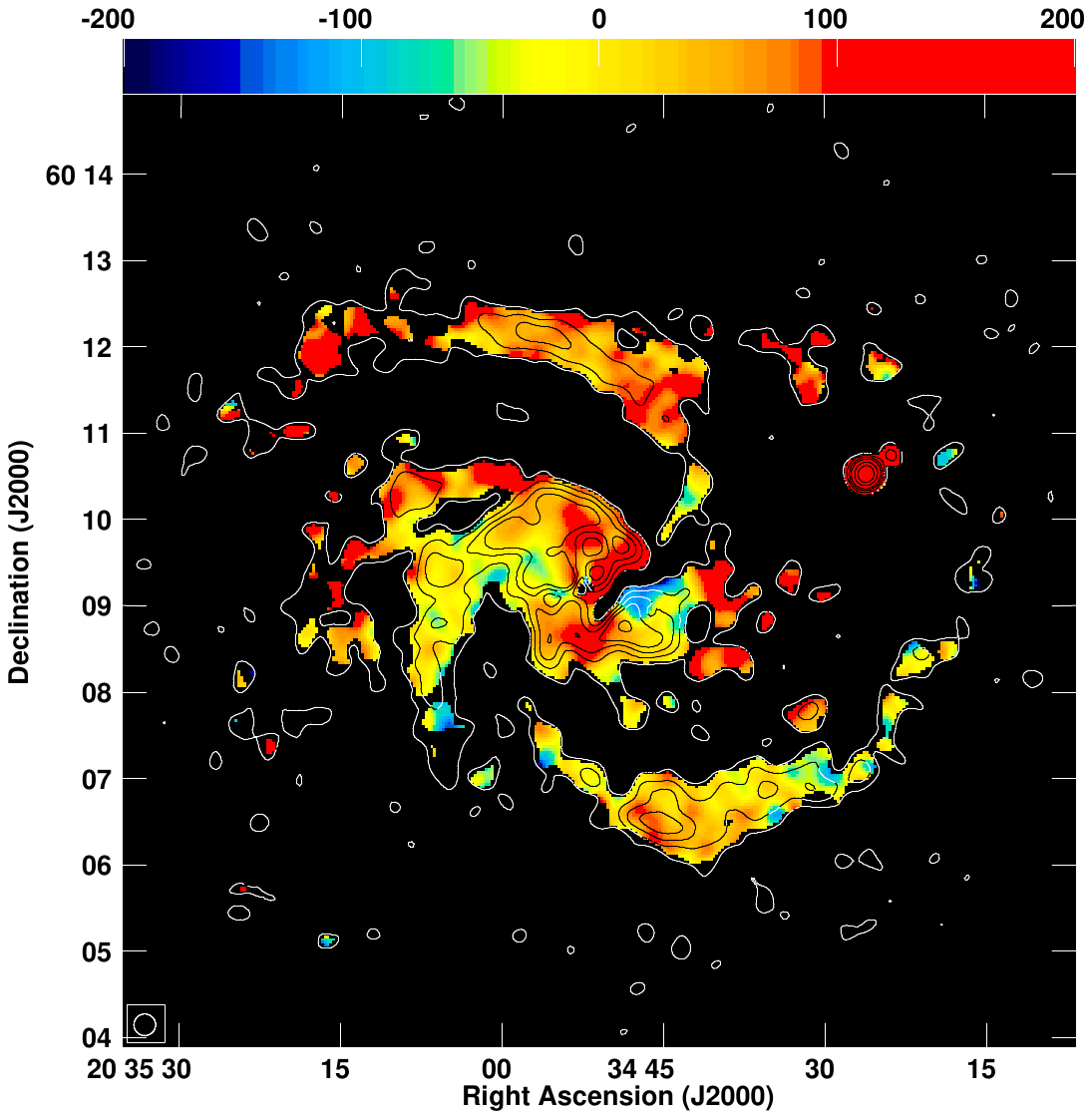}
    \end{minipage}

    \caption{Left panel: Magnetic field orientations overlaid on the H$\alpha$ emission of NGC6946 \citep{ferguson1998} and 6 cm polarized intensity contours (at 1,2,3,4,6 $\times$ 60 $\mu$Jy beam$^{-1}$ area) at a resolution of 15". Right panel: The Faraday rotation map of NGC6946 at 15" derived from 3 cm and 6 cm polarization data at pixels where the signal-to-noise in polarized intensity exceeds 6.5. The contours show the 6 cm polarized intensity at 1,2,3,4,6 $\times$ 60 $\mu$Jy beam$^{-1}$ area. A Galactic foreground RM of $+$40 rad m$^{-2}$ has been subtracted. RM values are predominantly positive (negative) in the northern (southern) arm with a large-scale sinusoidal azimuthal modulation. Detailed modeling of this modulation indicated the existence of an axisymmetric large-scale field superposed by higher order modes. Both figures are kindly reproduced by Rainer Beck based on his work in \cite{beck2007}.}
    \label{fig:ngc6946}
\end{figure}

In nearby galaxies, the most direct tracer of the plane-of-the-sky ordered magnetic field is polarized synchrotron emission at high frequencies as the emission suffers minimally from $\lambda^2$-dependent Faraday depolarization effects. Faraday rotation derived from multi-frequency diffuse polarized emission data provides the direction and strength of the line-of-sight component of the coherent (regular) magnetic field. While dust polarimetry of the SOFIA legacy program SALSA \citep{lm2022} has yielded beautiful magnetic maps of a dozen nearby galaxies, they lack field strength and field coherency information. To date, approximately 120 galaxies have had their galactic magnetic field properties derived mostly from narrowband diffuse radio polarization observations with limited angular resolution \citep{bw2013}. Figure \ref{fig:ngc6946} shows the magnetic field and the Faraday rotation maps of the well-studied NGC 6949 as an example of the typical polarization products. On the other hand, the coherent magnetic field along the line of sight can be directly revealed by the RM grid technique, where the Faraday rotation of a collection of background polarized radio sources is used to derive the magnetic field properties of the galaxy. At present, M31, the Large and Small Magellanic Clouds are the only three galaxies whose large angular extents enable the application of the RM grid \citep[e.g.][]{han1998,gaensler2005,mao2008}.

By recompiling a dozen of these archival datasets in a robust and consistent fashion, \cite{beck2019} re-derived several critical observed properties of the galactic magnetic field in nearby galaxies. The total magnetic field derived from the equipartition assumption \footnotemark[4]\footnotetext[4]{Caveats discussed in full details in \cite{sb2019}.} between cosmic rays and magnetic field in 12 nearby galaxies using 4.85 GHz polarization data has yielded a representative field strength of 13 $\mu$G \cite{beck2019}. The regular magnetic field has a typical strength of  1-2 $\mu$G. The random field (isotropic and anisotropic) dominates over the regular field, with an average B\textsubscript{reg}/B\textsubscript{tot} ratio of 0.13. The magnetic field strength correlates with the gas mass surface density which hints at an effective small-scale dynamo \citep{heesen2023b}. In terms of field geometry, galaxies exhibit a strong axisymmetric mode and rare large-scale field reversals (only the Milky Way has a confirmed large-scale field reversal \citep[e.g.,][see also discussions in Sun et al. in the same volume]{sun2008,vaneck2011,jf2012}). Only one candidate galaxy, NGC4666 could potentially have a reversal along the radial direction \citep{stein2019}. Halo fields have been observed to have much more complex structures with X-shape polarization signatures \citep{krause2020,irwin2024}, but the field coherency and symmetry in the halo is not well established in general. 

Observational measurements of the magnetic properties in the CGM of nearby galaxies are still lacking. The only available $\sim$ 3$\sigma$ detection of CGM magnetic fields comes from stacking of 183 nearby galaxies by \cite{heesen2023} using the LOFAR LoTSS DR2 RM catalog, which yielded an estimated CGM field strength of $\sim$0.5$\mu$G at the virial radius. No CGM RM grid studies have been carried out for individual galaxies.

\section{Pushing the envelope: magnetic field measurements at high redshifts with radio polarimetry}

Direct detection of the integrated polarized emission from distant galaxies is challenging: assuming a luminosity of a Milky-Way like galaxy and an integrated polarized fraction of 5\%, sub $\mu$Jy beam$^{-1}$ sensitivity is required to detect the polarized emission at  redshifts up to 0.5. Similarly, an angular resolution of better than 0.1" is needed to achieve physical resolution $<<$ 1 kpc for galaxies at z$>$0.5. These sensitivities and angular resolutions are challenging to meet with current instruments and will remain so in the foreseeable future. To date, no cosmologically distant star-forming galaxies have been detected in state-of-the-art deep ($\mu $Jy beam$^{-1}$ sensitivity) polarization surveys in the radio wavelengths (Ranchod et al. 2026, submitted). Therefore, even during the first phase of the SKA, we will mostly rely on back-illumination experiments to probe magnetic fields in distant galaxies. We note that recent ALMA polarimetry of a lensed galaxy revealed ordered galactic magnetic fields with spiral structures at z=2.3 \citep{geach2023,deroo2025}. However, from dust polarization data alone, it is possible not to distinguish between genuine large-scale coherent fields and mere-anisotropic random fields.

Faraday rotation studies of intervening galaxies seen against polarized background sources can reveal magnetic field properties of the foreground intervening galaxy. However, these studies are often hindered by (1) redshift dilution of the small expected RM signal; (2) isolation of Faraday rotation produced in the intervening galaxy from the intrinsic RM of the background radio source, and (3) the dominating Galactic foreground RM \citep{pandhi2025}. At small impact parameters, the use of gravitational lensing systems can directly address points (2) and (3) above, as the RM produced in the lensing galaxy can be obtained by differing the RM between the lensed images, effectively removing RM contributions from the Galactic foreground and the intrinsic background source properties\footnotemark[5]\footnotetext[5]{Under the following assumptions: (1) negligible RM fluctuations on arcsecond scales in the Milky Way foreground; and (2) negligible RM/ polarization variability intrinsic to the background source, or that any time-delay effects can be fully modeled and removed through long-term polarization monitoring}. \cite{Mao2017} and \cite{kovacs2025} measured magnetic fields through the disk and halo of two z=0.4 lensing galaxies. They found that at z=0.4, galaxies host disk-dominant coherent magnetic fields, which is consistent with those seen in present day galaxies. Substantial turbulent magnetic fields on small-scales ($<$ 50 pc) were revealed through modeling wavelength-dependent depolarization effects.  The lensing effect thus acts as a natural telescope, allowing us to peer into both the large-scale and small-scale magnetic fields in distant galaxies. These exciting initial findings warrant further investigations and expansion into different galaxy mass and redshifts.

\begin{figure}[t]
\includegraphics[width=0.99\textwidth]{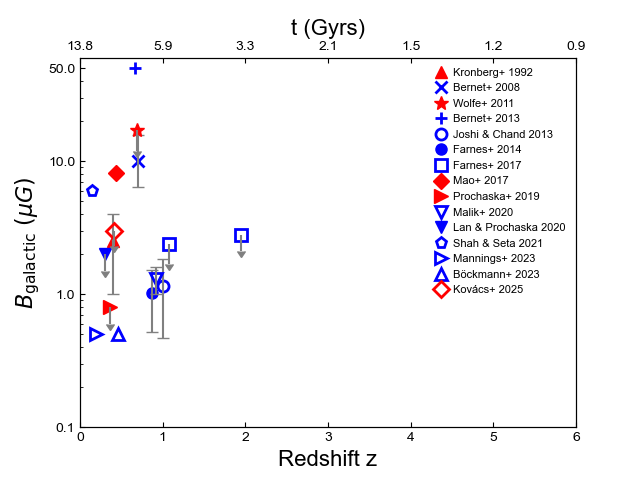}
\centering
\caption{A compilation of selected measurements of ISM and CGM magnetic fields beyond z=0. Blue symbols represent statistical measurements, while red symbols represent measurements in individual galaxies. Representative values from the following studies are included: \cite{kronberg1992,bernet2008,wolfe2011,bernet2013,jc2013,farnes2014,farnes2017,Mao2017,prochaska2019,malik2020,lan2020,shah2021,mannings2023,bockmann2023,kovacs2025}.} 
\label{fig:bvsz}
\end{figure}

At larger impact parameters, statistical studies have at best delivered marginal detections of magnetized gas associated with high-z galaxies, with most results being upper limits \citep[e.g.][]{kp1982,bernet2008,bernet2010,farnes2014,lan2020}. To minimize the impact of our inability to properly subtract the Galactic foreground RM, the sample could be limited in a small deep-drilling field. Recently, \cite{bockmann2023} utilized this approach based on MIGHTEE-POL polarization data of the COSMOS and XMM-LSS fields and found the CGM contributes an $|$RM$|$ of $\sim$ 5 rad m$^{-2}$ at 2.5$\sigma$ around z$\sim$0.5 galaxies. However, the magnetic field structure and turbulence level in the CGM remain unconstrained across z. In Figure \ref{fig:bvsz}, we have summarized the current status of ISM and CGM magnetic field measurements beyond the local Universe. The large apparent scatter can be attributed to estimates of different components of the magnetic field at different galacto-centric radii, obtained using a variety of tracers and methods across a wide range of galaxy types.

We note that dispersion measures (DM) and rotation measures of fast radio bursts have led to numerous single-sightline estimates of magnetic fields in the ISM and CGM of both nearby and distant galaxies \citep{prochaska2019,sherman2023,mannings2023}. Due to the unknown contributions from the host galaxy and local environment as well as our inability to accurately subtract the fluctuating Galactic RM foreground at the level needed to detect RMs produced in and around distant galaxies, the constraining power of these observations remains weak. However, future statistical studies of FRB DM and RM may yield more stringent limits \citep{kovacs2024,khrykin2025}.

\section {Current observational limitations}

Most of the archival diffuse polarized synchrotron data from galaxies were obtained prior to the advent of broadband receivers \citep{bw2013,bw2023}, hence most data available for the study of galactic magnetic fields are at discrete narrowband. RMs were usually computed using data at two wavelengths (3 cm and 6 cm), assuming a linear relation between the polarization angle and $\lambda^2$. Longer wavelength observations (11cm, 18cm, and 20cm) were often analyzed separately. The use of broadband data will prevent n$\pi$ ambiguity and allow modeling of RM and depolarization simultaneously across all bands without having to assume the dominant depolarization mechanism. Broadband polarization analyses including L and S bands (or even lower frequencies) would enable Faraday tomography to reveal the three-dimensional structure of the large-scale field and the turbulent properties of the small-scale field. 

Many of the historical datasets were from single-dish radio telescopes, with angular resolutions often exceeding 1', while the interferometric observations had angular resolutions that are rarely better than 10". For a typical local volume galaxy distance of 7 Mpc, this translates into a physical scale of $>$340 pc, far from the typical energy-injection scale of 50-100 pc in the ISM \citep{haverkorn2008,fletcher2011,ds2025}. As turbulence is unresolved in most of these previous observations, depolarization effects reduce the detectability of polarized emission, especially at lower frequencies, hindering us from obtaining a complete view of galactic magnetic fields. 

With the exception of CHANG-ES \citep{irwin2012,irwin2024} and WSRT SINGS \citep{heald2009,braun2010}, almost all galactic magnetic field studies in the nearby universe consist of a single galaxy, often selected because of their galaxy properties or expected/anticipated unique magnetic field features, with possibly vastly different observational setups. Only a large, homogeneous and unbiased dataset can facilitate rigorous comparison with theoretical predictions. We note that even non-detections of a large-scale coherent magnetic field are important for testing dynamo theory and its predictions.

The majority of existing extragalactic source RMs are narrowband data from the last century. Broadband observations of extragalactic radio sources (EGSs) will provide precise RMs without n$\pi$ ambiguity. By modeling the broadband complex polarized spectra, it would become possible to separate foreground from internal Faraday effects, thereby producing a cleaner RM grid. This is especially critical given the cumulative polarized source count density only increases with the survey detection limit to the power of -0.6 \citep{ro2014}. 

In order to successfully conduct an RM grid experiment, that is, to reliably recognize a simple large-scale ISM field pattern, we need to ensure sufficient ($>$ 50) sources behind the main body of a galaxy \citep{stepanov2008} for a statistically meaningful sample size of $>$40 galaxies. This requires a survey depth of at least 10 nJy beam$^{-1}$ to be reached at Band 2, which is well beyond the scope of the SKA. To reconstruct magnetic fields in external galaxies without prior knowledge of the field structure would require an even higher RM grid density (>1000 RMs per galaxy; \citep{stepanov2008}), which is only feasible for our closest neighbors, the Large and the Small Magellanic Clouds, in the SKA era. Therefore, RM grid studies of a large sample of nearby galaxies with the SKA will primarily focus on determining CGM magnetic fields owing to their much larger projected areas on the sky. 

At present, both the sample sizes for background polarized source- foreground galaxy pairs and for polarized lensing systems remain limited. The redshift range of distant galaxies probed are very limited as well, to mostly z$<$0.5.  

\section{Proposed SKA observations}

We propose four sets of key observations to be performed with AA4, targeting the ISM and CGM magnetic fields in both nearby and distant galaxies. 

{\bf (1) Mapping the diffuse polarized synchrotron emission in nearby galaxies:}
Ideally, we would require continuous frequency coverage from Bands 1, 2, 3, 4, and 5a to perform Faraday tomography and fully model the three-dimensional large-scale and small-scale magnetic fields. However, given the planned deployment of the frequency bands, we choose to prioritize observations in Band 5a because Faraday depolarization effects are minimal and we can achieve the highest precision in measuring magnetic field structures among the available bands. Intrinsic polarization angles and Faraday rotation can be derived from the broadband data within Band 5a. Even though Bands 1 and 2 will be available, we note that combined polarization analysis of Band 5a with Bands 1 and 2 in the absence of Bands 3 and 4 data is expected to be highly challenging. The large gap in frequency coverage would result in high sidelobe levels in Faraday depth space and subsequently significantly reduce our ability to constrain the magneto-ionic medium models through QU fitting \footnotemark[6]\footnotetext[6]{We note that at Band 5a, there may be contributions of anomalous microwave emission to the observed total intensity, thus appropriate correction is necessary prior to any polarization fraction analysis.}.  

With Band 5a, we expect an rms noise of 0.16$\mu$Jy beam$^{-1}$ in 50 hours, tapered to 4" resolution (a native resolution 0.14" with robust=0). This is sufficient to detect synchrotron emission from a region 100 pc in extent with a field strength of 9 $\mu$G (or 5$\mu$G field 1 kpc in extent) at better than the 7$\sigma$ level, representing an order-of-magnitude improvement in surface brightness sensitivity over existing surveys. The short baselines of SKA-Mid enable the recovery of polarized emission on kpc in nearby galaxies. A sample size of $\sim$ 50-100 galaxies would be desirable for proper statistical analysis. 

For a subset of nearby galaxies with edge-on orientation, targeted ultra-deep Band 1 or even SKA-LOW observations may reveal polarized emission associated with the steep-spectral-index synchrotron emission from aged cosmic ray electrons at large distances from the mid-plane. Thus, directly tracing magnetic fields outside the thin disk: in the thick disk, halo or even in the transition region to the CGM. 

{\bf (2) Probing the CGM magnetic fields of nearby galaxies using dense RM grids out to virial radius:} 
As the CGM field strength is low and expected to be in the sub $\mu$G regime, highly precise RM measurements would be needed. A 10$\sigma$ polarized detection in Band 1 would have an RM error of 0.4 rad m$^{-2}$ whereas a 10$\sigma$ polarized detection in Band 2 would have an RM error of 4 rad m$^{-2}$. The trade-off between polarized source density and the overall fractional polarization of sources (hence signal-to-noise ratio in polarization) and RM precision must be carefully considered before planning the final survey. To obtain at least 20 RMs through the CGM of $\sim$ 320 nearby galaxies, a sensitivity of 1 $\mu$Jy beam$^{-1}$ must be achieved, corresponding to at least 8 hours on-source time in Band 1 and 2 hours on-source time in Band 2.

{\bf (3) Deep, targeted follow-up polarization observations of individual lens candidates:}
We will select targets from lens search surveys (see McKean et al. in the same volume). As the surveys themselves likely lack the sensitivity to perform detailed polarization analysis and thus to obtain reliable RMs, we anticipate that the survey data are polarization calibrated so that they can at least provide candidate list for polarized systems. We would ideally prefer continuous frequency coverage in Bands 2, 3, 4, and 5a (Band 1 observations will not have sufficient angular resolution to resolve lensed images). Given the deployment plan, we would request Band 2 observations to provide precise RMs and Band 5a observations to best anchor the intrinsic polarization properties of the lensed radio source. 
Consider a minimum total flux density at Band 2 of 1 mJy beam$^{-1}$, we aim to detect 1\% polarization at the 10$\sigma$ level. This requires 1 $\mu$Jy beam$^{-1}$ sensitivity with robust=$-$1, an angular resolution of 0.43"x0.38", and 5 hours of on-source time per system in Band 2. Assuming a flat spectral index of $-$0.5, and a conservative lower bound on the polarization fraction at Band 5a, robust=0 (angular resolution of 0.14"x0.12"), an on-source time of 2 hours is needed. This is considered a modest investment of telescope time, provided that search surveys can already offer strong indications of whether the candidates are polarized. In addition to individual follow-up polarization observations, a large sample of polarized lensing systems identified in a high-resolution Band 5a survey has the potential to statistically detect magnetic fields in galaxies at even higher redshifts than those accessible with Band 2-only polarimetric data. 

{\bf (4) Statistical constraints of magnetic fields in CGM of distant galaxies using quasar absorption-line systems:}
We will devise deep polarimetric observations with a sky footprint matching on-going and future legacy spectroscopic surveys in the southern sky to maximize any cross-match potential and thus the sample size. Because the expected RM signal is small, Band 1 or/and Band 2 observations would be needed to provide the best RM precision. The survey area is likely to be on the order of a few hundred to 1000 sq degrees. Achieving a breakthrough on this front requires an RM grid at least an order of magnitude denser than any planned surveys. To reach a source density of $\sim$ 500 polarized sources per square deg (total of $\sim$ 5x10$^{5}$ polarized sources) requires a sensitivity of 0.1 $\mu$Jy beam$^{-1}$ corresponding to 2 hours per pointing in Band 2 with robust=0.

\section{Specific Science Focus}

{\bf Conducting a homogeneous galactic magnetic field census for a large sample of nearby galaxies}
From key science observation (1), we will be able to consistently derive various properties of galactic magnetic fields in a large number (50-100) of nearby galaxies and make conclusive observationally-based statements on the conditions required for galaxies to host large-scale magnetic fields, thereby directly testing dynamo theory. Comparisons of magnetic field properties to complementary HI observations will be needed to connect magnetic field properties with kinematics. Much can be learned on a per-galaxy basis (e.g., galaxy-specific dynamo models) as well as on a statistical basis using the entire sample. A combined analysis of polarization angles and RMs \citep{paul2025} can reveal dynamo modes in 3D, revealing how the spectrum of dynamo modes may depend on galaxy properties such as the number of spiral arms \citep{chamandy2013} and galaxy age. We will also be able to better assess the relative importance of possible magnetic arms generation mechanisms, including MHD waves \citep{Lou1998}. The properties of large-scale magnetic fields and their dynamic roles in different types of galaxies will be closely examined from dwarf galaxies \citep[e.g.][]{drzazga2016} to barred galaxies \citep[e.g.][]{beck2005} to elliptical galaxies \citep[e.g.][]{shah2021}. Detailed polarization studies of the Magellanic system \citep{gaensler2005,mao2008,mao2012,livingston2022,livingston2024} will inform us about the magnetic fields in and around our closest neighbors. 

Furthermore, detailed comparison with polarized dust emission in nearby galaxies to be mapped by the anticipated Atacama Large Aperture Submillimeter Telescope (AtLAST) will inform us of the interplay of magnetic fields in the multi-phase ISM \citep{liu2025}.

{\bf Resolving the magnetic field reversal tension}
Milky Way is the only galaxy with a confirmed large-scale magnetic field reversal along the Sagittarius arm \citep[e.g.][]{vaneck2011,korochkin2025}. One possible reason why they have not yet been confirmed in external galaxies is due to the insufficient angular resolution of current studies. Magnetic field reversals in spiral arms likely have limited radial and vertical extents, on the order of $\sim$ few hundred pc \citep{reid2019} or less. Resolving these reversals in a typical nearby galaxy at distance of 7 Mpc thus require an angular resolution of at least 5", which has not yet been achieved to date. The proposed key science observation (1) in combination with new techniques \citep{kurahara2021}, especially of edge-on galaxies, will ensure that the observational signatures of the field reversal are not washed out/depolarized and will inform us about the prevalence of these features and shed light on their origin.

{\bf Unraveling the elusive magnetic helicity}
Magnetic helicity and its conservation are important concepts in dynamo theory. However, strong observational evidence for helicity is still lacking, although large-scale magnetic field geometries consistent with helical fields have been reported in two external galaxies IC342 \citep{beck2015b} and NGC4217 \citep{stein2020}. The proposed key science observation (1) has the potential to reveal magnetic helicities for the first time. One likely manifestation would be observational signatures associated with opposite helicities of the plane-parallel fields in edge-on galaxies \citep{hs2014,bs2014}.

{\bf Quantifying the magnetized gas content in the CGM of nearby galaxies} 
Existing studies of the magnetized CGM in nearby galaxies rely on stacking approaches, and basic properties of the CGM field such as its extent, strength, and field geometry remain unknown. With key science observation (2), we can measure this in up to a hundred nearby galaxies individually and quantify CGM fields and its dependence on the underlying galaxy properties (e.g., with respect to outflow on a per-galaxy basis) for the first time \footnotemark[7]\footnotetext[7]{\cite{heesen2023} hints at a potential correlation between CGM magnetic fields and outflows along the minor axis based on stacking nearby galaxies.}. This information is critical for identifying the role of magnetic fields in the gas physics of the multiphase CGM.

{\bf Tracing the evolution of ISM magnetic fields over cosmic time} 
Key science observation (3) will unveil the first occurrences of large-scale magnetic fields in distant galaxies and place direct constraints on the time-scale of the large-scale dynamo. The much larger lens sample from the SKA will extend galactic magnetic field measurements to earlier cosmic epochs and a wider range of galaxy types. Comparisons with dynamo theory predictions and cosmological MHD simulations will effectively address the origin of large-scale galactic magnetic fields.

{\bf Investigating the redshift evolution of CGM magnetic fields}
Key science observation (4) will reveal the CGM magnetic fields in distant galaxies. The expected large sample size will allow binning in redshift space to characterize the redshift dependence of the CGM field. In addition, we will search for RM signatures that correlate with outflows, metal enrichment and the star formation rate history to test CGM magnetic field seeding mechanisms.

\section{Conclusions and Outlook}

In this chapter, we have presented our observational strategies with SKA AA4 to probe magnetic fields in and around nearby and distant galaxies. Carefully designed observations primarily in Bands 2 and 5a together will provide detailed magnetic field measurements in the ISM and CGM for a statistically substantial sample of nearby and distant galaxies for the first time. These key polarimetric observations will significantly advance our knowledge of the magnetic fields in the ISM and CGM, enabling one to characterize the evolution of this elusive non-thermal component of the baryonic cycle as a function of redshift. 

We would like to emphasize that beyond the first generation receivers, polarimetric observations in Bands 3 and 4 are critical in all of the outlined science cases above as $\sim$ 1.7 - 5 GHz is the optimal polarimetric observing frequencies for galactic turbulence characterized by Faraday dispersion in the range of $\sim$ 10 -100 rad m$^{-2}$ \citep{ab2011}. Continuous frequency coverage bridging Band 2 and 5a data is indispensable for comprehensive modeling of the three-dimensional large-scale and small-scale magnetic fields in external galaxies \citep{heald2015}. 

Looking beyond the current baseline design into the SKA 2 Era, the envisioned improvement in sensitivity and angular resolution will allow one to directly probe the magnetic fields using polarized synchrotron emission in galactic regions of weak emission (e.g., at large galactocentric radii, or in extreme turbulent regions). The combination of the order of magnitude increase in sensitivity and increase in FoV will increase the survey speed, boosting the RM grid density by a factor $>>$ 4 (within a given observing time) than AA4, enabling more nearby external galaxies to have an RM grid constructed for determining their ISM fields, as well as a denser CGM RM grid in nearby and distant galaxies. The expected 20-fold improvement in the resolving power in the GHz frequency range will significantly increase the number of strong lensing candidates and hence expand the sample size of high redshift galaxies in which ISM fields can be probed by the lensing method, with accessibility to galaxy masses down to dwarf galaxy and to even higher redshifts. It is without a doubt that the full SKA will revolutionize our understanding of the origin of magnetic fields in and around galaxies.

\section{Acknowledgments}
Andrea Bracco acknowledges financial support from the INAF initiative ``IAF Astronomy Fellowships in Italy'' (grant name MEGASKAT). Rikuto~Omae was supported by JSPS KAKENHI Grant Number 24KJ0111. Theresa Wiegert acknowledges financial support from  the grant CEX2021-001131-S  funded by MICIU/AEI/ 10.13039/501100011033, from the coordination of the participation in SKA-SPAIN, funded by the Ministry of Science, Innovation and Universities (MICIU). Theresa Wiegert acknowledges financial support from the grant PID2021-123930OB-C21 and PID2024-155817OB-I00 funded by MICIU/AEI/ 10.13039/501100011033 and by ERDF/EU.

\bibliographystyle{abbrvnat-maxbibnames4}
\bibliography{chapter}

\end{document}